# Phase Diagram of QCD at Finite Temperatures with Wilson Fermions

Y. Iwasaki[a]

[a]Institute of Physics, University of Tsukuba, Tsukuba, Ibaraki 305, Japan

Phase diagram of QCD with Wilson fermions for various numbers of flavors $N_F$ is discussed. Our simulations mainly performed on a lattice with the temporal size $N_t = 4$ indicate the following: The chiral phase transition is of first order when $3 \leq N_F \leq 6$, while it is continuous when $N_F = 2$. For the realistic case of massless u and d quarks and the strange quark with $m_q = 150$ MeV, the phase transition is first order. The sharp transition in the intermediate mass region for $N_F = 2$ at $N_t = 4$ observed by the MILC group disappears when an RG improvement is made for the pure gauge action.

## 1. Introduction

In this article I discuss the finite temperature phase diagram of QCD with Wilson fermions. In this formulation chiral symmetry is explicitly broken by the Wilson term even for vanishing bare mass. The lack of chiral symmetry causes much conceptual and technical difficulties in numerical simulations and in physics interpretation of data. Therefore we start our discussion with the problem of how one defines the quark mass and the chiral limit in sect. 2. In sect. 3 we examine two fundamental problems: whether the chiral limit of the finite temperature transition exists at all, and if it does exist, what is the order of the chiral transition. The strategy we take to investigate these problems is to carry out simulations along the line of the critical hopping parameter ("on-$K_c$" simulation). With this method we have identified the chiral transition points and the orders of the transition for the cases of $N_F = 2$, 3 and 6. Then a more realistic case of two massless quarks and one light quark is discussed.

Recently the MILC group reported an unexpected finding of a first-order transition at the intermediate quark mass region for the $N_F = 2$ case. Another strange phenomenon is that the quark mass calculated at high temperatures does not agree with that at low temperatures when $\beta < 5.5$. These points are discussed and possible resolution with an RG improved action will be given in sect. 4. Finally conclusions will be given in sect. 5.

## 2. Definition of quark mass

The chiral property of the Wilson fermion action was carefully investigated through Ward identities for axial vector currents by Bochicchio et al.[1]. Independently we proposed[2] to define the quark mass by

$$2m_q < 0\,|\,P\,|\,\pi > = -m_\pi < 0\,|\,A_4\,|\,\pi >$$

where P is the pseudoscalar density and $A_4$ the fourth component of the local axial vector current. We first showed in quenched QCD at zero temperature that the pion mass vanishes at the point which is almost identical to that where the quark mass vanishes. Maiani and Martinelli[3] also showed this for a quark mass similarly defined through the axial-vector Ward identity.

It has been shown that the value of the quark mass does not depend on whether the system is in the high or the low temperature phase through simulations at $\beta = 5.85$ in the quenched QCD [4] and at $\beta = 5.5$ for the $N_F = 2$ case[5]

Fig.1 shows numerical results of the critical value of the hopping parameter $K_c$ for $N_F = 2$. Result for $\beta \lesssim 5.0$ are obtained by two methods; either from the vanishing point of the extrapolated $m_\pi^2$ or from that of the extrapolated $m_q$ in the confining phase [6–11]. The data at $\beta=6.0$ and 10.0 are new, which are determined from $m_q=0$ in the deconfining phase. A characteristic feature in this figure is a sharp change of $K_c$ around $\beta \approx 5.0$ signalling a transition from the strong coupling region to the weak coupling re-gion.

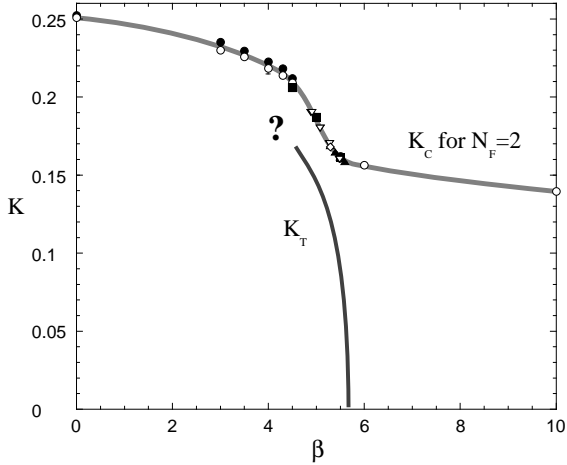

Figure 1. The line of $K_c$ for $N_F = 2$.

## 3. Chiral transition

In Fig.1 a schematic line is drawn for the line of transition $K_T(\beta)$ separating the high temperature regime from the low temperature regime for a fixed temporal lattice size $N_t$. Whether the $K_T$ line crosses the $K_c$ line is not trivial. As was first noticed by Fukugita et al.[12], the $K_T$ line creeps deep into the strong coupling region.

To solve this problem, we take the strategy of performing simulations on the line $K_c(\beta)$ starting from a $\beta$ with the system in the high temperature phase and reducing $\beta$. We call this method "on-$K_c$" simulations. The number of iteration $N_{\text{inv}}$ needed for the quark matrix inversion provides a good indicator to discriminate the two phases[13] because there are zero modes around $K_c$ in the low temperature phase, while none exits in the high temperature phase. Combining this with measurements of Polyakov loop, plaquette and hadron screening masses, we identify the crossing point $\beta_{ct}$ of the $K_T(\beta)$ line with the $K_c(\beta)$ line. We also check that the crossing point thus determined is consistent with an extrapolation of the line $K_T(\beta)$ toward the chiral limit. From the behavior of physical quantities toward $\beta_{ct}$, we are also able to determine the order of the chiral transition by this method.

The values of $K_c(\beta)$ are slightly different (at most of the order of 0.01) depending the definition of the chiral limit. They also depend on $N_F$. We find that the difference due to $N_F$ is of the same order of magnitude as that due to the difference of definition. We take the largest (farthest) $K_c$ in order to see the existence of the crossing point since this is the most stringent condition.

### 3.1. $N_F = 2$

Several groups[12,7,15,5,9–11] investigated the problem of the finite temperature transition for the $N_F = 2$ case. The results for $K_T(\beta)$ for $N_t = 4$ and 6 together with those for $K_c(\beta)$ are plotted in fig.2. The curves are for guiding eyes.

We identified the crossing point $\beta_{ct}$ by the "on-$K_c$" simulation[9] to be at $\beta_{ct} \sim 3.9 - 4.0$ for $N_t = 4$ and at $\beta_{ct} \sim 4.0 - 4.2$ for $N_t = 6$. The existence of the crossing point is important, because otherwise we can not reach the chiral limit in the confining phase[12].

The change of $N_{inv}$ with $\beta$ is gradual and we have not observed two-state signals. Furthermore the pion screening mass squared gradually decreases towards zero as the chiral transition point is approached[9]. See also Fig. 1 in [14]. These results imply that the transition is continuous for $N_F = 2$. This is in sharp contrast with the case of $N_F = 3$ and 6 where clear two-state signals are observed.

For $N_t = 18$ with the spatial size $18^2 \times 24$, we find that the transition is at $\beta_{ct} \sim 4.5 - 5.0$[9]. Although the spatial size is not large enough, this result suggests that the shift of $\beta_{ct}$ with $N_t$ is small.

### 3.2. $N_F = 3$ [14]

The phase diagram for $N_F = 3$ with $N_t = 4$ together with the $K_T$ line for $N_F = 2$ is shown in Fig.3. We found clear two-state signals at the crossing point located at $\beta_{ct} = 3.0$ by the "on-$K_c$" simulation[9] as shown in Fig. 4.

An interesting question is to determine the critical value of the quark mass $m_q^{\text{crit}}$ up to which the first order phase transition persists.

We observed clear two state signals at $\beta = 4.0$,

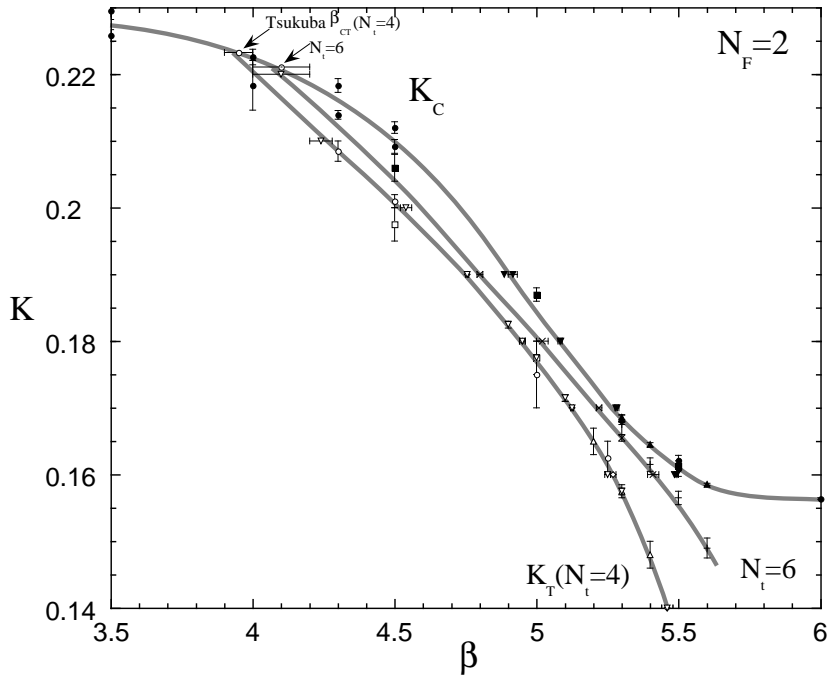

Figure 2. The transition/crossover lines $K_T$ for $N_t = 4$ and 6 together with the $K_c$ line for $N_F = 2$.

4.5 and 4.7, while for $\beta = 5.0$ and 5.5 no such signals have been seen. The value of the quark mass $m_q a$ at the transition of $\beta = 4.7$ equals 0.175(2). Using $a^{-1} \sim 0.8$ GeV obtained from hadron spectrum calculations[14], we then obtain a bound on the critical quark mass $m_q^{\rm crit} \gtrsim 140$ MeV or equivalently $(m_\pi/m_\rho)^{\rm crit} \gtrsim 0.873(6)$

We note that these values are much larger than those for the staggered fermion case for which the critical mass is from 10 to 40 MeV and $(m_\pi/m_\rho)^{\rm crit} \simeq 0.42 - 0.58$. ($m_q a = 0.025 - 0.075$[16,18]. We have used the value of $a^{-1} = 500$MeV at $\beta = 5.1 - 5.2$ for $N_F=4$[19], because the data of $a$ for $N_F = 3$ are not available.)

### 3.3. $N_F = 6$ and $N_F \geq 7$

We have also examined the $N_F = 6$ case and identified the crossing point at $\beta_{ct} = 0.3$[9], where we see clear two-state signals (see Fig. 4). The small value of $\beta_{ct}$ is consistent with our finding that the $K_T$ line does not cross the $K_c$ line at finite $\beta$ when $N_F \geq 7$[13].

It should be noted that the results that the chiral transitions for $N_F = 3$ and 6 are of first order, while it is continuous for $N_F = 2$, are consistent with the prediction based on universality[17].

### 3.4. $N_F = 2 + 1$ [14]

Now let us discuss a more realistic case of massless u and d quarks and a strange quark. When the mass of the strange quark changes from zero to infinity the order of the phase transition should change from first to continuous at some critical mass. A very important issue is whether the first order transition persists up to the physical strange quark mass. Our strategy to investigate this problem is similar to the "on-$K_c$" simulation; we set $K = K_c$ for u and d quarks and fix the difference between the strange quark mass and the u and d quark masses to some value, and make simulations changing the value of $\beta$.

We have studied two cases corresponding to the strange quark mass of about 150 MeV and 400 MeV in terms of the lattice spacing estimated from the $\rho$ meson mass (See Fig.3 in [14].). It should be noted that the physical strange quark mass determined from $m_\phi = 1020 MeV$ turned out to be $m_s \sim 150 MeV$[14] for our definition of the quark mass.

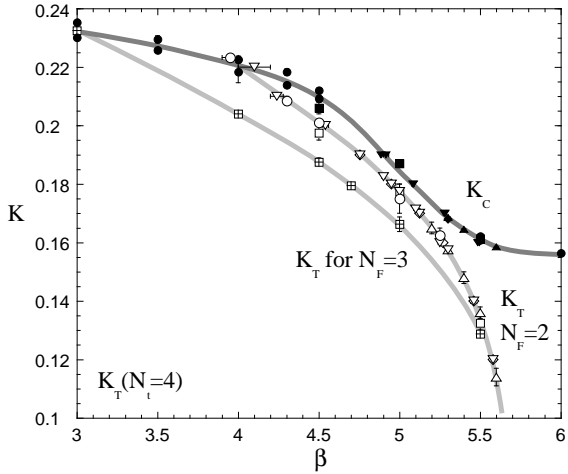

Figure 3. The transition/crossover line $K_T$ for $N_F = 3$ together with that for $N_F = 2$ and the $K_c$ line.

For both cases we observed clear two state signals, at $\beta = 3.5$ for 150 MeV and at $\beta = 3.9$ for 400 MeV as shown in Fig. 4. Thus our results suggest that the transition for the physical strange quark mass is of first order.

This result is apparently in conflict with the result for staggered quarks by the Columbia group[18] that no transition occurs at $m_u a = m_d a = 0.025$, $m_s a = 0.1$ ($m_u = m_d \simeq 12$ MeV, $m_s \simeq 50$ MeV using the value of $a$ from [19]), which suggests that a first-order phase transition does not occur for the physical u,d and s quarks. Note in this connection that the results for $N_F = 2+1$ with Wilson and staggered quarks are both consistent with the results for degenerate $N_F = 3$ cases, respectively, that $m_q^{\rm crit} \gtrsim 140$ MeV for Wilson quarks and $m_q^{\rm crit} \simeq 10 - 40$ MeV for staggered quarks. ( It is satisfied that for $m_s \lesssim m_q^{\rm crit}$ the transition is of first-order, as it should be.) However, consistency between Wilson and staggered quarks is an issue which should be investigated in future.

## 4. Crossover behavior at $\beta \approx 5.0$ for $N_F = 2$

### 4.1. MILC results

The MILC group extensively investigated the crossover region at $\beta \approx 5.0$ for the $N_F = 2$ case and found unexpected phenomena[10,11]. For $N_t = 4$, the transition is smooth both for heavy and light quarks. However, in the range of intermediate masses corresponding to $\beta \sim 5.0$, the transition is very sharp. This is completely opposite to what was supposed to be realized.

In Fig.5, the MILC data[10] at $\beta = 5.0$ together with ours[9] are plotted. The change of the Polyakov loop is sharp and $m_\pi^2$ has a cusp. Further the quark mass in the high temperature phase does not agree with that in the low temperature phase and shows peculiar behavior. This strange behavior was also noticed by us in various cases (see, for example, Fig. 1 in [13]).

The strange behavior of the quark mass for $\beta < 5.5$ implies that the high temperature phase in this region is far from the continuum limit. On the other hand, the chiral transition even on a $N_t = 18$ lattice with the spatial size $18^2 \times 24$ occurs at $\beta < 5.5$. Therefore in order to obtain quantitative predictions for physical quantities in the high temperature phase we have to perform simulations at $N_t \gtrsim 18$.

Furthermore, for $N_t = 6$, the MILC group observed clear two state signals in the range of intermediate mass ($K = 0.17$ and $\beta = 5.22$, $K = 0.18$ and $\beta = 5.02$, and $K = 0.19$ and $\beta = 4.8$). They suggested the possibility that the transition at $K = 0.19$ is a bulk transition based on their numerical results.

Now let us carefully look at Fig.2 again. The $K_T$ line for $N_t = 6$ starts from the $K_c$ line at $\beta \approx 4.0 - 4.2$, and initially deviates from it. We observe, however, that the $K_T$ line approaches the $K_c$ line at $\beta \approx 4.8 - 5.2$ and $K \approx 0.17 - 0.19$. It is very interesting to note that this is exactly the region where two state signals are observed. I believe that the appearance of first order phase transitions in the intermediate mass region is closely related with this unusual relation between the $K_T$ line and the $K_c$ line. It is plausible that the first order phase transition and the strange behavior of the quark mass illustrated in Fig.5 are lattice artifacts.

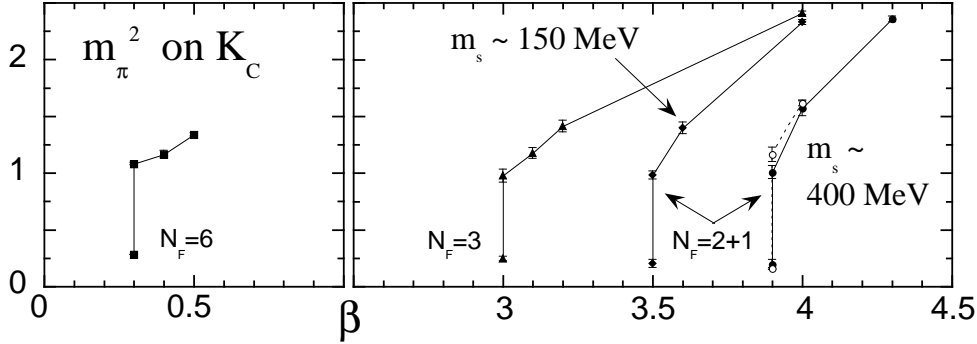

Figure 4. The pion screening mass squared on the $K_c$ line for $N_F =$ 6, 3 and 2+1. The filled symbols are for a $8^2 \times 10 \times 4$ lattice, while the open ones are for a $12^3$ lattice.

### 4.2. Possible resolutions with an RG improved action [20]

We have suggested that the problems encountered by the MILC group at intermediate quark masses are lattice artifacts. There are two possible approaches to avoid this situation: 1) Perform simulations on a lattice with large $N_t$. However, we have to make simulations on a lattice with $N_t \gtrsim 18$, which is very difficult. 2) Use an RG improved action for simulations, for which the approach to the continuum limit should be smoother.

In connection with the second possibility, we proposed about ten years ago an RG improved action for pure gauge action of the form[21]

$$S_{gauge}^{IM} = 1/g^2(c_0 \Sigma W(1 \times 1) + c_1 \Sigma W(1 \times 2))$$

with $c_1 = -0.331, c_0 = 1 - 8c_1$, where $W(1 \times 1)$ and $W(1 \times 2)$ are the Wilson loops of sizes $1 \times 1$ and $1 \times 2$, and the sum is over all oriented loops.

With this improved action we calculated the string tension and hadron spectrum, and investigated the stability of instantons and the $U(1)$ problem in quenched QCD [2,22].

The results we interpret as improvements may be summarized as follows. 1) The ratio $\Lambda_{\overline{MS}}/\Lambda_{IM} = 0.488$ is of order unity in contrast to $\Lambda_{\overline{MS}}/\Lambda_{Standard} = 28.81$, which implies that the bare coupling is already an improved coupling. 2) The approach of the plaquette to the perturbative value is much faster than the case of the standard one-plaquette action. 3) A first-order phase transition found in pure SU(5) gauge theory with the standard action disappears with the improved action. 4) Concerning stability of instantons, the improved action is located on the boundary separating the stable and unstable regions in the parameter space of the action.

With these experiences it is worthwhile to use the RG improved action for full QCD. There is also a possibility to improve the quark action. However we take the Wilson action for the quark action as a first step, because we believe that the effect of improvement of the gauge sector is much more significant than that of the quark sector by the following reasoning. After integration over fermion variables, the action may be written as a sum over various types of Wilson loops. The coefficients thus obtained are numerically much smaller at simulation points compared with $c_1 = -0.331$ for the improved pure gauge action.

In Fig.5 we compare the results for the standard action at $\beta = 5.0$ with those for the improved action at $\beta = 2.0$. The inverse lattice spacings are identical; 0.99(6) GeV for the standard action[7] and 1.01(4) GeV for the improved action. In the case of the improved action, there is no sudden change of the quark mass in the high temperature region and its value agrees well with that in the low temperature region. The change of the Polyakov loop is smooth, and the pion mass

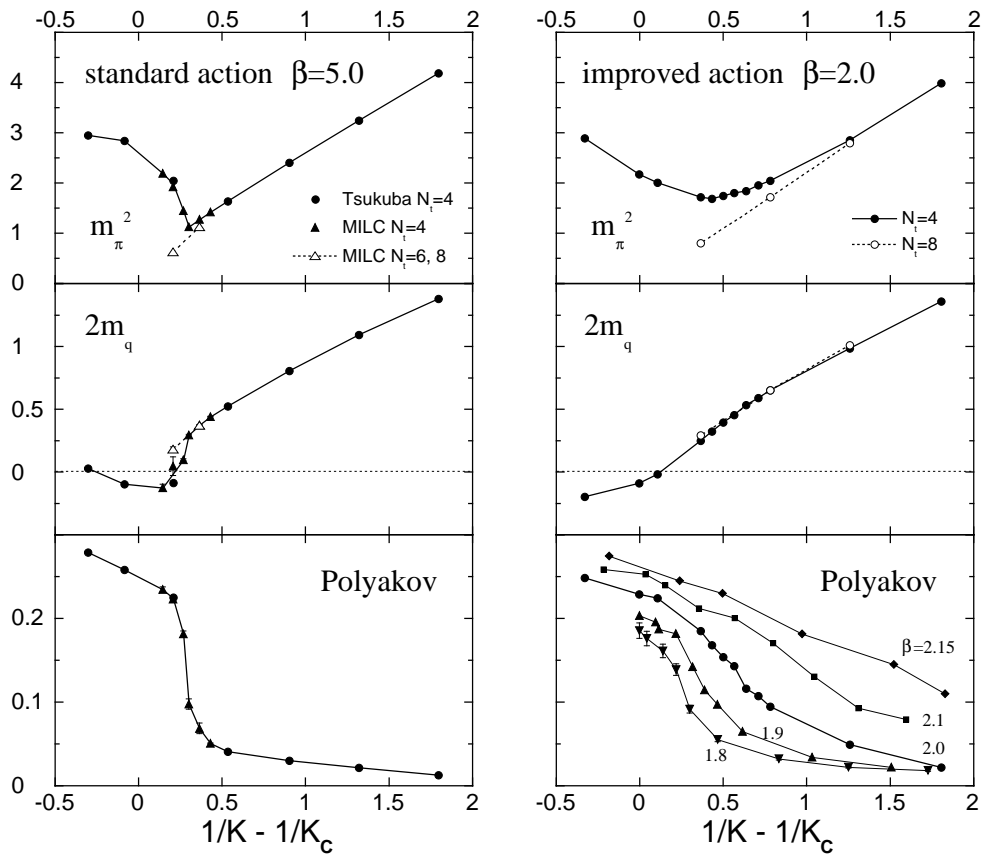

Figure 5. The data on $N_t = 4$ (filled symbols) and $N_t = 6$ or 8 (open symbols) lattice at $\beta = 5.0$ with the standard action[9,10] and at $\beta = 2.0$ with the improved action. The Polyakov loops are shown in a range from $\beta = 1.8$ to 2.15 for the improved action. The $K_c$ is from $m_\pi^2 = 0$ at $N_t = 8$.

squared has no cusp.

Preliminary results in the range $\beta = 1.6 - 2.15$ indicate that the transition is smooth in the intermediate mass region ($\beta \sim 2.0$) and becomes sharper toward the chiral transition which is located at $\beta \sim 1.5 - 1.6$. (See Polyakov loops in Fig.5.) Therefore we conclude that the sharp transition in the intermediate mass range disappears with the RG improved action at $N_t = 4$. This suggests that the first order phase transition observed by the MILC group at $N_t = 6$ disappears with this action. To confirm this, we are planning to extend runs to a wider range of parameters, in particular, on a lattice with $N_t = 6$.

## 5. Conclusions

Investigation of the phase diagram of QCD with Wilson fermions is conceptually and technically difficult due to the lack of chiral symmetry. Therefore we first have to grasp the gross feature of the phase diagram. In this aspect the research along the boundary of the phase diagram is important. The nontrivial boundary, the $K_c$ line, can be determined by the vanishing point of the quark mass defined through the axial vector Ward identity. Then simulations along the $K_c$ line as well as in the strong coupling limit give useful information for the phase structure. Main results obtained in this way may be summarized as follows. For the degenerate case, the chiral phase transition is of first order when $3 \leq N_F \leq 6$, while it is continuous when $N_F = 2$. For the realistic case of massless u and d quarks and the strange quark with $m_q = 150$ MeV, the phase transition is first order.

When $N_F \geq 7$, the $K_t$ line does not cross the $K_c$ line at finite $\beta$. In the strong coupling limit, when the quark mass is lighter than a critical mass, the chiral symmetry is not spontaneously broken and the quark is not confined[13]. This is the zero temperature phase structure. I hope we can discuss the phase diagram at zero temperature in detail elsewhere.

Recent work by the MILC group has revealed unexpected phenomena, which occur not along the boundary of the phase diagram. To understand the phenomena we have to investigate the phase diagram in a wider parameter space of the action. We find that when an RG improvement is made for the pure gauge action, both the sharp transition in the intermediate mass region for $N_F = 2$ at $N_t = 4$ and peculiar behavior of the quark mass in the high temperature phase for $\beta < 5.5$ disappear.

Of course, we have to confirm the conclusions with larger $N_t$ and need much more work to derive quantitative estimates of various physical quantities.

I thank members of QCDPAX group for their collaboration. I am grateful to C. DeTar for providing the MILC data and A. Ukawa for useful discussions and for valuable suggestions on the manuscript. This work is in part supported by the Grant-in-Aid of Ministry of Education, Science and Culture (No.06NP0601).